\documentclass[floatfix,prb,aps,twocolumn,showpacs]{revtex4}

\usepackage{epsfig}
\usepackage{color}
\newcommand{\chunks}{r}
\newcommand{\dw}{W_m}
\begin{document}

\title{Stripe formation in doped Hubbard ladders}

\author{G. Hager}
\author{G. Wellein}
\affiliation{Regionales Rechenzentrum Erlangen, Martensstra\ss e 1,
D-91058, Erlangen, Germany}
\author{E.~Jeckelmann}
\affiliation{Institut f\"{u}r Physik, Johannes
Gutenberg-Universit\"{a}t, D-55099 Mainz, Germany}
\author{H. Fehske}
\affiliation{Institut f\"{u}r Physik, Ernst-Moritz-Arndt-Universit\"{a}t, 
Greifswald, D-17489 Greifswald, Germany}

\date{\today}

\begin{abstract}
  We investigate the formation of stripes in $7\chunks \times 6$
  Hubbard ladders with $4\chunks$ holes doped away from half filling
  using the density-matrix renormalization group (DMRG) method.  A
  parallelized code allows us to keep enough density-matrix
  eigenstates (up to $m=8000$) and to study sufficiently large systems
  (with up to $7\chunks = 21$ rungs) to extrapolate the stripe
  amplitude to the limits of vanishing DMRG truncation error and
  infinitely long ladders.  Our work gives strong evidence that
  stripes exist in the ground state for strong coupling ($U=12t$) but
  that the structures found in the hole density at weaker coupling
  ($U=3t$) are an artifact of the DMRG approach.
\end{abstract}


\maketitle

Two-dimensional lattice models for correlated electrons are often used
to describe the properties of layered cuprate compounds.~\cite{Dagotto}
Despite numerous studies there is an ongoing controversy about the 
existence of stripes in these systems.
In a hole doped system a stripe is a domain wall ordering of 
holes and spins. The wall is made of a narrow hole-rich region.
The spins are antiferromagnetically ordered between the walls
and are correlated with a $\pi$ phase shift across a wall. 
The formation of stripes in the ground state has been 
demonstrated numerically for the $t-J$ model on (narrow)
ladders~\cite{WS03,WSLadders} using the density-matrix
renormalization group (DMRG) method.~\cite{dmrg}
For square lattices, however, the presence of stripes
remains controversial~\cite{Prelov93,WS2D,Manousakis,Sorella} because
a reliable investigation of the ground state in the 
thermodynamic limit is not possible with the methods 
currently available.

Recently, attention has turned to the two-dimensional 
Hubbard model with a local electron-electron repulsion $U$
and an electron hopping term $t$.
For $U=0$ this model describes a Fermi gas, which obviously
has no stripes in the ground state.
Moreover, no instability toward the formation of stripes
has been found in the weak-coupling limit $U \ll t$
using renormalization group techniques.~\cite{Lin97}  
In the 
strong-coupling limit $U \gg t$, however,
the Hubbard model can be mapped onto a
$t-J$ model with $J = 4t^2/U \ll t$,
which does have stripes in the ground state,
at least on narrow ladders with $J \approx 0.35t$.~\cite{WS03,WSLadders}
Therefore, investigating the formation of
stripes in the Hubbard model at finite coupling $U/t$
could significantly improve our understanding 
of these structures.
Moreover, such investigations should reveal the true
capability of the various methods used
to study stripes much better than calculations for the
$t-J$ model alone.

An early DMRG investigation of 3-leg Hubbard ladders~\cite{Bonca00}
found that stripes formed in the ground state only for 
$U \geq 6t$.
In a recent DMRG calculation White and Scalapino~\cite{WS03} 
have shown that a narrow stripe appears in the ground state of 6-leg 
Hubbard 
ladders (more precisely, $7 \times 6$-site clusters) for $U \geq 6t$.
For weaker couplings the hole and spin densities show structures
which are interpreted as a broad stripe.
In both works, however, no finite-size scaling has been performed
and the amplitude of the hole density 
modulation has not been investigated systematically as a function
of DMRG truncation errors. 
Thus it is not clear if the observed structures are really
the signature of a striped ground state 
in the limit of infinitely long ladders; 
they could be finite-size effects or an artifact of the DMRG method.  

Here, we report on a DMRG investigation of stripes in
6-leg Hubbard ladders with up to 28 rungs. 
Keeping up to $m=8000$ density-matrix eigenstates
the amplitude of the hole density modulation can be
extrapolated to the limit of vanishing DMRG truncation errors
for systems with up to $21$ rungs.
This allows us to perform a reliable finite-size scaling analysis
of the hole density modulation. 
Calculations for systems of that size are made possible
by a parallelized shared-memory DMRG code which runs 
efficiently on current supercomputer architectures.~\cite{Hager03}
We show that the stripes found by White and Scalapino~\cite{WS03}
are stable
in the limit of an infinitely long ladder for strong coupling
$U=12t$. For weak coupling ($U=3t$), however, the hole 
density fluctuations found in Ref.~\onlinecite{WS03} are an
artifact of truncation errors and boundary conditions.  

The two-dimensional Hubbard model on a $R\times L$ ladder
is defined by the Hamilton operator
\begin{eqnarray}
\hat{H} 
&=& -t \sum_{x,y,\sigma} 
\left( \hat{c}_{x,y,\sigma}^{\dag}\hat{c}_{x,y+1,\sigma}^{\phantom{\dag}}
 + \hat{c}_{x,y,\sigma}^{\dag}\hat{c}_{x+1,y,\sigma}^{\phantom{\dag}}
 + \text{h.c.} \right )
\nonumber \\
&&+ U \sum_{x,y} \hat{n}_{x,y,\uparrow}
\hat{n}_{x,y,\downarrow}  \;,
\label{Hamiltonian}
\end{eqnarray}
where $x=1,\dots,R$ is the rung index
and $y=1,\dots,L$ is the leg index,
$\hat{c}^{\dag}_{x,y,\sigma}$ and 
$\hat{c}_{x,y,\sigma}$ are creation and annihilation operators for
an electron with spin $\sigma=\uparrow,\downarrow$ at site $(x,y)$,
and $\hat{n}_{x,y,\sigma}=
\hat{c}^{\dag}_{x,y,\sigma}\hat{c}^{\phantom{\dag}}_{x,y,\sigma}$ is the 
corresponding density operator. 
Here we exclusively consider the Hubbard model on 6-leg ladders ($L=6$) with
$R=7\chunks$ rungs for $\chunks=1,\dots,4$. Cylindrical boundary conditions
were used (closed in the rung [$y$] direction 
and open in the leg [$x$] direction), because they
are the most favorable ones for DMRG simulations.
Moreover, open boundaries break the translational
invariance of the system, allowing spin and charge structures to appear
as local density variations in a finite ladder.
If periodic boundary conditions were used, one would have to analyze 
correlation functions to detect stripes in finite ladders.
Since we are interested in the ground state of the hole-doped regime,
we consider a system with $N = 4\chunks$ holes doped in the half-filled
band, corresponding to $RL -N = 38\chunks$ electrons.
The average hole density is $n = N/RL = 4/42 \approx 0.095$
for all cases, as in Ref.~\onlinecite{WS03}.

We employ a recently developed parallelized DMRG code~\cite{Hager03}
to determine the ground state properties of this Hubbard model.
Our DMRG program implements the standard finite-system
algorithm~\cite{dmrg} for two-dimensional lattices.
Parallelization of the most time-consuming tasks
allows us to carry out calculations of
unprecedented magnitude.
For the calculations presented here
we have kept up to $m=8000$ density-matrix eigenstates
per block for systems with up to $R\times L= 168$ sites. 
This requires up to four weeks walltime and
100 GBytes of memory per run on eight processors of an IBM p690 node.
For comparison, it takes about 6 hours 
to reproduce
the results of Ref.~\onlinecite{WS03} for $7\times 6$ clusters
with $m=3600$\@.
DMRG calculations have already been performed for the Hubbard
model on larger systems (square lattices or ladders) than our 
$28 \times 6$-site clusters but for a significantly smaller
number of density matrix eigenstates 
($m \leq 2000$).~\cite{WS2D,Nishimoto02}
The computational cost of these simulations was at least an order
of magnitude lower than in the present work.

For our 6-leg Hubbard ladders,
the standard DMRG method yields the ground state energies
and various expectation values for the ground state
of the system investigated.
Here, we focus on the hole density
\begin{equation}
h(x,y) = 1-\langle  
\label{hole}
\hat{n}_{x,y,\uparrow}+\hat{n}_{x,y,\downarrow} \rangle
\end{equation}
and the staggered spin density
\begin{equation}
s(x,y) = (-1)^{x+y}\left \langle 
\hat{n}_{x,y,\uparrow}-\hat{n}_{x,y,\downarrow} \right \rangle
 \;\;,
\label{spin}
\end{equation}
where $\langle \dots \rangle$ represents the (DMRG)
ground state expectation value. 
In the first few lattice sweeps of our DMRG calculations
or for a small number $m$ of density-matrix eigenstates per block
$(m \alt 1000)$, the DMRG wavefunction reaches a `metastable' 
state,~\cite{WS2D,Bonca00} which
depends essentially on the initial conditions, i.e.,
on the detail of the method used to construct the lattice in the
first sweep (for more detail about the DMRG method, 
see Ref.~\onlinecite{dmrg}).
The hole and staggered spin densities  
show irregular fluctuations in both the rung and the leg directions
at that point of the DMRG calculation.
\begin{figure}
\centerline{%
\epsfig{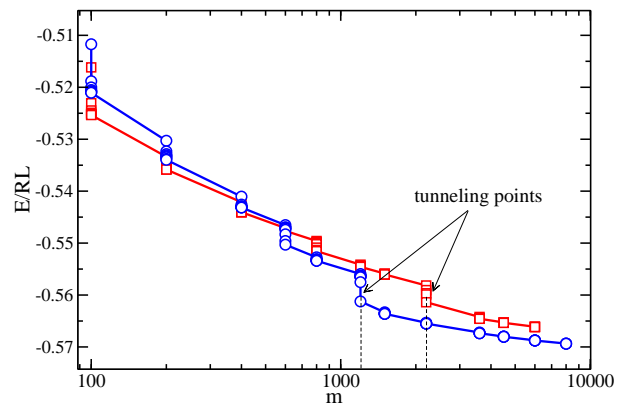}}
\caption{(Color online) DMRG ground state energy per
site versus $m$ for the $21\times 6$ ladder at $U=12t$,
using (circles) and not using (squares) block reflection
in the middle of the lattice. Every data point corresponds to one sweep
of the DMRG procedure.
\label{fig:eoverm}}
\end{figure}

For all system sizes and coupling strengths investigated,
the DMRG wavefunction ``tunnels'' to a stable state after several 
sweeps and for sufficiently large $m$, as reported in Ref.~\onlinecite{WS03}
for $7 \times 6$ clusters. We have observed that the larger the 
system length $R$, the larger must $m$ be
to reach the stable state, ranging e.g.\ from
$m\approx 600$ at $R=7$ to $m\approx 2200$ at $R=21$ in the
$U=12t$ case\@. This state is then essentially 
independent of the initial conditions, but it is nevertheless
essential to make $m$ as large as feasible in order to
get sufficient data for a reliable extrapolation of
observables (see below)\@. 
The tunneling occurs for smaller $m$ when it is possible
to utilize the block reflection technique (see Fig.~\ref{fig:eoverm})\@.
Note, however, that the combination of spin and charge density fluctuations
can easily break the symmetry between left and right DMRG blocks and that 
the block reflection technique should not be used in that case
as it can lead to incorrect results.
\begin{figure}
\centerline{%
\epsfig{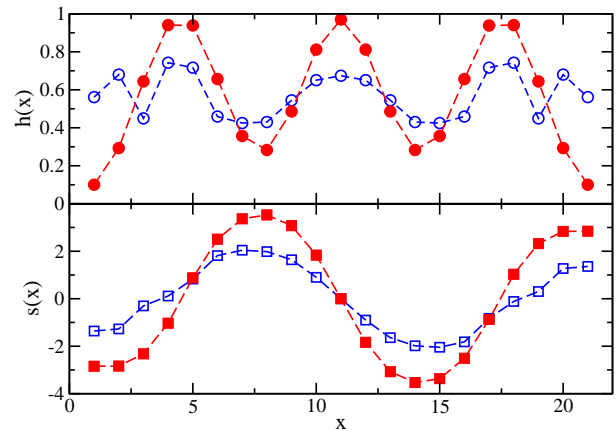}}
\caption{(Color online) Hole (circle) and staggered spin (square) densities
in the leg [$x$] direction calculated  
on a $21\times 6$ Hubbard ladder with 12 holes for 
$U=3t$ (open symbols) and $U=12t$ (solid symbols). 
\label{fig:profiles}}
\end{figure}
For the $7 \times 6$
case, it is interesting to note that the 
transition occurs at a significantly smaller $m$ with our
algorithm than in Ref.~\onlinecite{WS03}\@. This may also be attributed
to a more favorable choice of initial conditions.
Our results for large $m$ are in full agreement
with those presented in Ref.~\onlinecite{WS03}.
The ``tunneling'' is marked by a sharp drop in energy,
while the spin and hole densities become more regular. 
In particular, the hole density and the staggered
spin density are almost constant in the rung direction.
The stability of this DMRG `ground state' is demonstrated by the
systematic behavior of the energy (for all system sizes) and 
expectation values (for systems with up to 21 rungs)
as a function of the discarded weight (see
the discussion below).
Some of our results for $28 \times 6$ ladders are
inconclusive because an insufficient number of sweeps
has been performed for the charge and spin density profiles
to reach convergence.  

On the $7 \chunks \times 6$ ladders with $4\chunks$ holes investigated here,
$\chunks$ stripes with 4 holes each appear in the DMRG ground state.
These stripes are clearly seen in the hole density modulation
in the leg direction
\begin{equation}
h(x) = \sum_{y=1}^{W} h(x,y) \; ,
\end{equation}
which is shown in Fig.~\ref{fig:profiles} for a $21\times6$ ladder
with $U=12t$.
In the same figure, one sees that the staggered spin 
density in the leg direction
\begin{equation}
s(x) = \sum_{y=1}^{W} s(x,y) \; 
\end{equation}
is finite and changes sign exactly where the hole density $h(x)$ 
is maximal.
Therefore, the specific features of stripes are clearly observed
in the DMRG ground state densities.
Note, however, that the finite staggered
spin density is an artifact of our DMRG method~\cite{WS03},
which does not use the full spin symmetry. 
In the true ground state of a finite ladder
one expects $s(x,y)=0$. 
In Fig.~\ref{fig:profiles} the results for 
$U=3t$ appear qualitatively similar to those obtained for $U=12t$
although   
the amplitudes of the density fluctuations are 
smaller for the weaker coupling. 
Nevertheless, one notices that
the hole and spin density profiles for $U=3t$ 
are less regular than for $U=12t$. 

To make a quantitative analysis of these structures we 
have carried out a systematic spectral analysis of the
hole and staggered spin densities. 
The spectral transforms are defined as 
\begin{equation}
F(k_x,k_y) = \sqrt{\frac{2}{L(R+1)}} 
\sum_{x,y} \sin(k_x x) e^{i k_y y} f(x,y)
\end{equation}
with $k_x = z_x\pi/(R+1)$ for integers $z_x=1,\dots,R$ and
$k_y = 2\pi z_y/L$ for integers $-L/2 < z_y \leq L/2$.
Here $f(x,y)$ and $F(k_x,k_y)$ represent either the hole density 
$h(x,y)$ and its transform $H(k_x,k_y)$ or
the staggered spin density $s(x,y)$ and its transform $S(k_x,k_y)$.
The transformation in the rung direction (with periodic 
boundary conditions) is the usual Fourier transform.
In the leg direction (with open boundary conditions)
we use an expansion in particle-in-the-box eigenstates because 
this is a natural basis for a finite open system.
In the infinite-ladder limit $R \rightarrow \infty$ this transformation
becomes equivalent to the standard Fourier transformation.
As the systems considered have a reflection symmetry
($x \rightarrow R+1-x$), the hole spectral transform $H(k_x,k_y)$
always vanishes for even integers $z_x$ while
the spin spectral transform $S(k_x,k_y)$ vanishes
for odd $z_x$ if $R$ is odd and for even $z_x$ if $R$ is even. 
Moreover, in the converged DMRG ground state we observed
uniform behavior of $h(x,y)$ and $s(x,y)$ along the rungs.
This implies that the spectral weight is concentrated
at $k_y=0$ for both densities.

\begin{figure}
\centerline{%
\epsfig{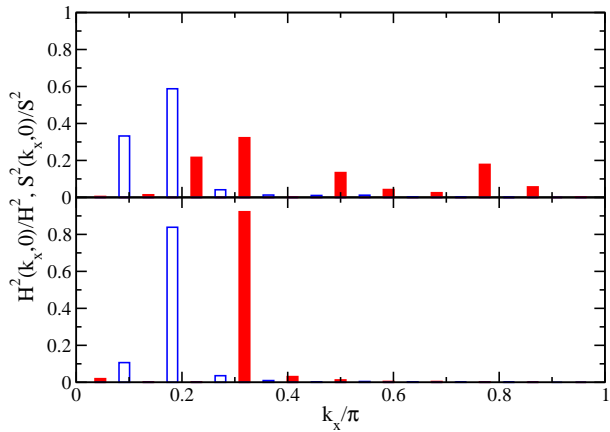}}
\caption{(Color online) Normalized
power spectrum of the hole (solid bar) and staggered spin 
(open bar) 
densities presented in Fig.~\ref{fig:profiles}
for a $21 \times 6$ Hubbard ladder with
$U=3t$ (upper panel) and $U=12t$ (lower panel). 
\label{fig:transforms}}
\end{figure}

In Fig.~\ref{fig:transforms} we show the power spectrum
(squared norm of the spectral transforms)
of the hole and staggered spin densities presented in 
Fig.~\ref{fig:profiles}.
In both cases, the power spectrum has been normalized by its
total weight 
\begin{equation}
F^2 = \sum_{k_x,k_y}  |F(k_x,k_y)|^2  \;,
\end{equation}
which we denote $H^2$ and $S^2$ for the hole and spin power spectrum,
respectively.
For $U=12t$ one sees that both hole and spin power spectra 
have a single strong peak containing most of the spectral weight
(92 \% and 84\%, respectively).
For $U=3t$, however, we observe a
broad distribution without a clearly dominant mode $k_x$
in the hole power spectrum. 
We find similar results for other ladder lengths $R \leq 21$.
For $R=28$ the power spectra are mostly inconclusive because
of the non-convergence of the hole and spin densities mentioned
previously.

For $U=12t$ the observed positions of the dominant peaks in the 
hole and spin spectral transforms for $r\in\{1,2,3\}$ can be extrapolated to
the $R=\infty$ limit, yielding
\begin{equation}
{k_x^{\mathrm H}\over\pi} = 
{2\chunks+1\over R+1}~\stackrel{R\to\infty}{\longrightarrow} 
~ {2\over 7}
\end{equation}
and 
\begin{equation}
{k_x^{\mathrm S}\over\pi} = 
{\chunks+1\over R+1}~\stackrel{R\to\infty}{\longrightarrow} 
~{1\over 7}~,
\end{equation}
respectively, which agrees perfectly with the expected values
corresponding to one stripe every seven rungs in an infinitely long
ladder.
For $U=3t$, the position $k_x$ of the maximum in the spectral transforms
is not always well defined (for instance,
it changes with the number $m$ of density 
matrix eigenstates kept even for large $m$) and thus
a quantitative analysis of $k_x$ for $R\rightarrow \infty$ is not possible.

All DMRG calculations suffer from truncation errors which are 
reduced by increasing the number~$m$ of retained density matrix 
eigenstates 
(for more details, see Refs.~\onlinecite{dmrg})\@.  
The error in the ground state energy is proportional to the
discarded weight $\dw$ which is defined as the
total weight of the discarded density-matrix eigenstates:
\begin{equation}
\dw = \sum_{i=m+1}^d w_i\,.
\end{equation}
Here, $d$ is the dimension of the density matrix and
$w_i$ is its $i$th eigenvalue.
Thus one can achieve a greater accuracy 
and obtain an estimate of the error with
a linear extrapolation of the DMRG energy to the limit of vanishing
truncation errors $\dw \rightarrow 0$ (for an example of this
extrapolation, see Ref.~\onlinecite{Bonca00})\@.
Using this technique we have found that the error in the ground 
state energy per site is typically about $4 \times 10^{-3}t$
($U=12t$) and $7 \times 10^{-3}t$ ($U=3t$)
for the largest number of density-matrix eigenstates kept
($m=8000$ and $m=6000$, respectively) and
$R \leq 28$. 
Consequently, the error in the total energy 
is of the order of $t$ for the largest system (168 sites)\@.
The energy separation between
ground state and the lowest excited states, which is of the order
of a small fraction of $t$, is thus
significantly smaller than the error in the total energy.
Therefore, the DMRG wavefunctions obtained in our calculations
are not
accurate descriptions of the true ground states.
Although the DMRG wavefunctions converge systematically to
the true ground states (as shown by the linear
scaling of the energy with $\dw$), 
for $m \leq 8000$
they still have significant overlaps
with other eigenstates. 
Expectation values calculated with such a DMRG
wavefunction (i.e., for a given $m$) could thus be quite inaccurate.
In order to get reliable results we will in the following
carefully analyze
the scaling of observables with increasing $m$.
\begin{figure}
\centerline{%
\epsfig{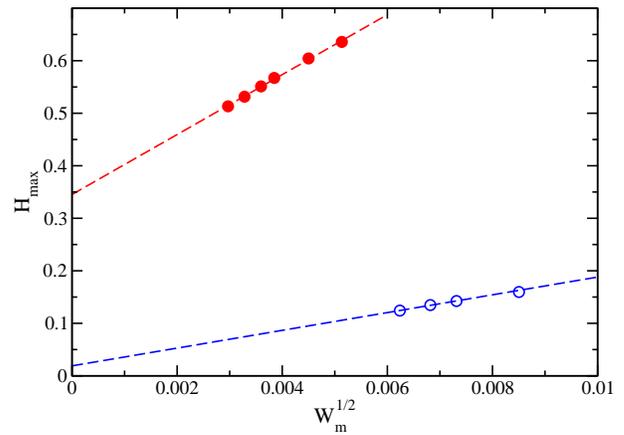}}
\caption{(Color online) Maximum $H_{\text{max}}$
of $|H(k_x,0)|$ in a $21 \times 6$ system
with $U=12t$ (solid circle) and $U=3t$ (open circle)
as a function of the square root
of the discarded weight $\dw$ for various numbers
of density-matrix eigenstates $1800 \leq m \leq 8000$.
Dashed lines are linear fits.
\label{fig:truncation}}
\end{figure}

If a variational ground-state wavefunction, as used in the DMRG
algorithm, is known up to an error of $\varepsilon$, the corresponding
error in the energy is of the order of $\varepsilon^2$\@.  Other
observables, whose operators are nondiagonal in the energy basis,
converge only with $\varepsilon$\@.  For DMRG we know that
$\varepsilon^2\propto \dw$ (cf. Ref.~\onlinecite{Bonca00}), thus 
expectation values of operators are polynomials 
of $\sqrt{\dw}$ for small discarded weights $\dw$\@.
For the maximum $H_{\text{max}}$ of the absolute hole spectral transform
$|H(k_x,0)|$  we find a linear scaling with $\sqrt{\dw}$
(see Fig.~\ref{fig:truncation})\@. 
Such a scaling has already been found for other density modulation
amplitudes.~\cite{Jeckelmann} 
This allows us to extrapolate $H_{\text{max}}$ to the limit of vanishing
truncation errors.
We note that $H_{\text{max}}$ decreases with decreasing $\dw$.
This corresponds to a diminution of the stripe amplitude with
increasing $m$ (i.e., decreasing $\dw$) for sufficiently large
$m \agt 2000$. This diminution can be seen directly in the hole density 
profile $h(x)$ (for instance, in Fig.~3b of Ref.~\onlinecite{WS03})\@.
For $U=12t$ the extrapolated values of $H_{\text{max}}$ are clearly
finite as shown in Fig.~\ref{fig:truncation} for a
$21 \times 6$ ladder.
Thus we conclude that the hole density fluctuations
found on finite ladders
are not an artifact of DMRG truncation errors but a feature
of the true ground state for $U=12t$.   
For $U=3t$, $H_{\text{max}}$ extrapolates to very small values
as $\dw \rightarrow 0$.
Typically, the range of $\sqrt{\dw}$ over which we observe
a linear behavior in $\sqrt{\dw}$ is smaller than the smallest value
of $\sqrt{\dw}$ used in the extrapolation. This can be seen
for the example shown in Fig.~\ref{fig:truncation}. 
The uncertainty in the extrapolated $H_{\text{max}}$
is thus larger than its value for $U=3t$\@.
Therefore, the hole density fluctuations could be the result
of DMRG truncation errors and the true ground state could be uniform
in that case, i.e., $H_{\text{max}} = 0$ for $\dw \rightarrow 0$.

\begin{figure}
\centerline{%
\epsfig{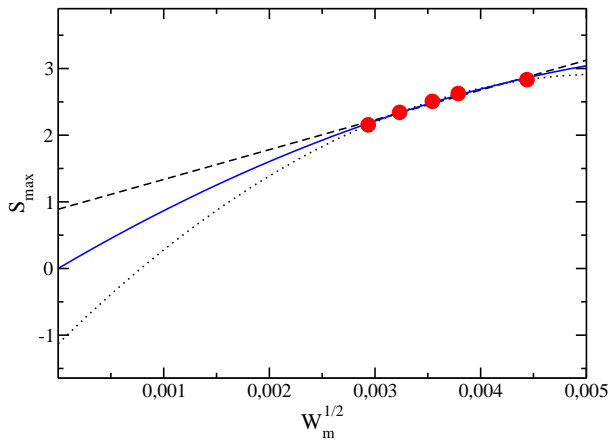}}
\caption{(Color online) Maximum $S_\mathrm{max}$ of $|S(k_x,0)|$
for the $7 \times 6$ system.  
Dashed line: linear fit, dotted line: quadratic
fit, solid line: quadratic fit with constraint $S_\mathrm{max}(0)=0$\@. 
\label{fig:spinex}}
\end{figure}

Extrapolating the maximum $S_{\text{max}}$ of the absolute spin spectral transform
$|S(k_x,0)|$ to the limit $\dw \rightarrow 0$ turns out to be more
difficult than extrapolating $H_{\text{max}}$\@.
Contrary to $H_{\text{max}}$, $S_{\text{max}}$ has not reached
an asymptotic regime (as a function of $\dw$)  for the largest
number $m$ of density matrix eigenstates we have used.
This difference is probably due to the smaller energy scale
of spin excitations compared to that of charge excitations, 
resulting in a DMRG ground state which describes the charge properties
of the true ground state correctly but not its spin properties.
Nevertheless, for the smallest ($7 \times 6$)  ladder an 
extrapolation of $S_\mathrm{max}$
to $\dw \rightarrow 0$ using linear and quadratic fits 
suggests that $S_{\text{max}}$ 
vanishes for $\dw \rightarrow 0$
and thus that the true ground state has no spin density fluctuations,
$s(x,y)=0$, as expected (see 
Fig.~\ref{fig:spinex}). 
\begin{figure}[t]
\centerline{%
\epsfig{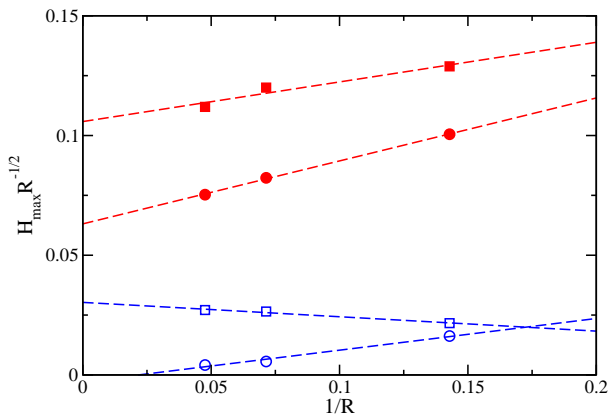}}
\caption{(Color online)
Amplitude $H_{\text{max}}/\sqrt{R} \propto h_0$ of the hole density
modulation for a fixed number ($6000 \leq m \leq 8000$) 
of density-matrix eigenstates (square) 
and extrapolated to the limit $\dw \rightarrow 0$ 
(circle) 
as a function of the inverse ladder length $1/R$ 
for $U=12t$ (solid symbols) and $U=3t$ (open symbols).
Dashed lines are linear fits.
\label{fig:size}}
\end{figure}
The artificial breaking of the spin symmetry is similar for
all couplings $U$, indicating that it does not affect the formation of
stripes (i.e., the existence of a hole density modulation),
which is a strongly $U$-dependent phenomenon as shown here.

Typically, the discarded weight $\dw$ is about $10^{-5}$ or smaller
for $m=8000$.
Although this appears to be a small value, the above discussion
shows that errors in 
the ground state energy and $H_{\text{max}}$ are still quite large
for that number $m$.
This confirms that the absolute value of the discarded weight $\dw$
alone does not give a reliable estimate for errors on
physical quantities.

A ladder with a periodic array of stripes has 
a modulation of the hole density (charge density wave)
\begin{equation}
h(x) = h_0 \sin(k_x^Hx) \; ,
\end{equation}
which corresponds to  
\begin{equation}
H_{\text{max}} = |H(k_x^H,0)| = \sqrt{\frac{(R+1)L}{2}} h_0  \; .
\end{equation}
If the amplitude $h_0$ of the hole density modulation is finite
in the limit of an infinitely long ladder ($R \rightarrow \infty$),
the maximum $H_{\text{max}}$ of the spectral transform must diverge
as $\sqrt{R}$ for $R \rightarrow \infty$.
In Fig.~\ref{fig:size} we show the finite-size scaling of
$H_{\text{max}}/\sqrt{R} \sim h_0$ as a function of the inverse ladder
length for $U=3t$ and $U=12t$.
The values of $H_{\text{max}}/\sqrt{R}$ obtained for a fixed number
$m$ of density matrix eigenstates converge to finite values for
$R \rightarrow \infty$, suggesting the existence of stripes
in this limit for both couplings $U$\@.
After extrapolation to the limit of vanishing DMRG truncation errors,
however, $H_{\text{max}}/\sqrt{R}$ seems to vanish for large $R$ in the
case $U=3t$
while it still converges to a finite value for $U=12t$
(see Fig.~\ref{fig:size})\@.
This indicates that stripes exist in the ground state of 
infinitely long
ladders for sufficiently strong coupling such as $U=12t$ but that the
hole and spin structures found in finite ladders for
weak couplings such as $U=3t$ are artifacts of open boundaries
and DMRG truncation errors.
It has already been observed in other problems~\cite{Jeckelmann} 
that DMRG truncation
errors can result in an artificial broken symmetry ground state
after extrapolation to infinite system sizes, while extrapolating first
to the limit of vanishing truncation errors restores the true ground state 
symmetry in the thermodynamic limit.

In conclusion, we have investigated the formation of stripes in
6-leg Hubbard ladders at a hole doping of 9.5\%.
Using a parallelized DMRG code we have been able to
determine the amplitude
of the hole density modulation in the limits of vanishing
DMRG truncation errors and infinitely long ladders.
Our results show that stripes exist in the ground state of
these systems for strong but not for weak couplings.

\acknowledgments

We gratefully acknowledge a fruitful collaboration with the Zuse
Institut Berlin (ZIB) who granted exclusive access to computational
resources on their HLRN complex. This work was partly supported by the
Competence Network for Scientific High Performance Computing in
Bavaria (KONWIHR)\@.

\end{document}